\documentclass[fleqn,10pt,twocolumn]{SICE24}

\newcommand{\figref}[1]{Fig.~\ref{#1}}
\newcommand{\equref}[1]{Eq.~(\ref{#1})}

\usepackage{siunitx}
\usepackage{enumitem}
\usepackage{amsmath} 
\usepackage{float} 

\title{Detecting $\sim$10 mK Face Temperature Change Based on \\Lock-in Thermography Referencing Heartbeat}

\author{Nanami Kotani${}^{1\dagger}$ and Yasuaki Monnai${}^{2}$}
\speaker{Nanami Kotani}

\affils{${}^{1}$Graduate School of Information Science and Technology, The University of Tokyo, Tokyo, Japan\\
(nanami.kotani@star.rcast.u-tokyo.ac.jp)\\
${}^{2}$Research Center for Advanced Science and Technology, The University of Tokyo, Tokyo, Japan\\
(monnai@star.rcast.u-tokyo.ac.jp)\\
}

\abstract{
Infrared thermography, which has widely spread particularly during the COVID-19 period, has been effectively used for research on health monitoring and emotion estimation. Nevertheless, detecting minute temperature changes with thermography is challenging as it is disturbed by not only noise but also outside temperature surrounding the object. In this study, we demonstrate detecting face temperature variation by implementing lock-in thermography using heartbeat signals as a reference. It allows us to detect minute temperature changes, as low as $\sim$10 mK, on the forehead with a commercially available thermal camera. The proposed approach enables stable measurement of body temperature variation, showing potential for non-contact emotion estimation.
}


\begin{document}

\maketitle


\section{Introduction}
Infrared thermography enables remote temperature measurement 
applicable to multiple people regardless of lighting conditions \cite{Chan2006JTM}.
The use of thermography has also been explored beyond temperature measurement for diagnosing illnesses \cite{Clark2007Allergy,DAlessandro2024AS}
 and estimating emotions \cite{Jody2015ER,SalazarLopez2015CC,CruzAlbarran2017IPT,Egger2019ENTCS} based on the fact that human skin temperature changes via blood circulation due to heartbeat, respiration, sweating, and muscular activity \cite{Ioannou2014Psy}. Its non-contact nature offers advantages over methods like electromyography or electroencephalography, which require physical contact. 

There have been many important studies that have applied thermography to measure blood flow or visualize blood vessels.
For example, a spectral filtering approach has established a conversion between temperature signal and blood flow in the hand and feet \cite{Sagaidachnyi2017PM}. Another study has demonstrated blood vessel imaging in the forearm and back of the hand from thermography after applying pressure to the upper arm and arterial occlusion for 5 minutes \cite{Liu2012TF}.
Other studies have used lock-in thermography to visualize veins in the hand \cite{Bouzida2008SPIE}
, in which a change of the skin temperature around the veins synchronous with the pressure application to the arm is detected. In \cite{Bouzida2009JTB}, veins on the back of the hand have been visualized by using the fact that the skin temperature above the veins rises when the hand is cooled to maintain the temperature balance.

While many studies have been conducted for hands and arms, 
detecting a change in the face temperature remains challenging.
The face temperature can indeed be used to measure a variety of human physiological indicators such as respiration rate \cite{Jagadev2020IPT}, blinking, and blood flow \cite{Sagaidachnyi2017PM} with the potential for emotion estimation.
When estimating emotion from the face temperature, regions around the eyes, cheeks, forehead, tip of the nose, and mouth are often selected as typical ROIs (Region of Interests) \cite{Jody2015ER}. However, it is difficult to accurately select ROIs on the forehead and nasal tip vessels and non-reproducible, and the variation in the definition of ROIs across different studies could lead to inconsistency \cite{SalazarLopez2015CC}.

To tackle this challenge, we propose a method to implement lock-in thermography for the face temperature using a commercially available thermal camera by using the heartbeat as a reference signal. By significantly enhancing the signal-to-noise ratio, we experimentally demonstrate to observe minute face temperature as low as $\sim$10 mK, synchronous to the heartbeat. The result will be important for emotion estimation as well as medical applications.


\section{Method} \label{sec:Method}
\begin{figure}[t]
    \centering
    \includegraphics[width=\linewidth]{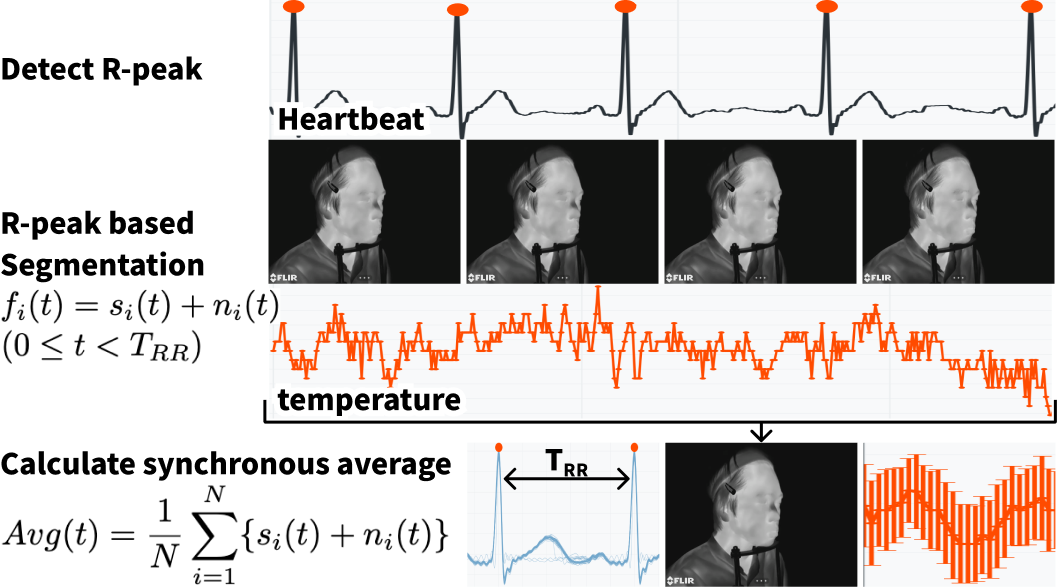}
    \caption{Overview of the proposed lock-in thermography using the ECG as a reference timing signal.}
    \label{fig:method}
\end{figure}

An overview of the proposed method is illustrated in \figref{fig:method}. In this study, lock-in thermography is implemented to detect a minute change of the face temperature, as small as $\simeq$ 10 mK, which will be important for sensing mental and physiological information. Synchronous averaging of the thermal images is calculated by using the electrocardiogram (ECG) as a reference timing signal. In this way, a temperature change synchronous to the heartbeat is enhanced owing to coherent addition while a random noise is reduced owing to incoherent addition.
Specifically, the following procedure was employed for our experiment.

\begin{enumerate}[leftmargin=14pt]
    \item  Timing Detection: Peak search of the ECG is performed to detect the R-peaks, the maximum voltage appearing in each cycle based on ventricular excitement, using the Christov's algorithm \cite{Christov2004BEO}.
    \item  Temporal Axis Scaling: The temporal axis of the quasi-periodic R-peaks is locally scaled to be periodic. The same scaling is applied to the frame timing of thermal images.
    \item  Averaging: Additive averaging of the periodically scaled thermal images is calculated. 
\end{enumerate}


After recording $N$ cycles of the heartbeat, we scale the temporal axis based on ECG to obtain $N$ temperature profiles at a constant interval applicable to synchronous averaging. Let the time-scaled temperature during one cycle at a certain location of the thermal camera be $f_i (t)\mbox{ }(i=1,2,\ldots,N)$. It is expressed as the sum of the true temperature $s_i (t)$ and the noise $n_i (t)$ as 
\begin{align}
    f_i (t) &= s_i (t) + n_i (t) ~~ (0 \leq t < T_{RR}),
    \label{eq:frameExtraction}
\end{align}
where $T_{RR}$ is the scaled period of the RR interval. $T_{RR}$ changes with each lock-in because the RR interval varies with the individual and the individual's condition at the time of measurement. The RR interval is the time from one R-peak to the next, and normal values are approximately 0.6 to 1.2 seconds.
The averaging is thus calculated as
\begin{align}
    Avg(t) &= \frac{1}{N} \sum_{i=1}^N \{ s_i (t) + n_i (t) \}.\label{eq:synchronousAdditiveAverage}
\end{align}
In practice, we keep recording the thermal images for a longer period. We then slide this averaging window to capture the temporal shift of averaged signals.

The standard error $SE$ is derived for the synchronous additive averaging. The standard error $SE$ can be expressed as follows, with standard deviation $SD$ where $\bar{f}$ denotes the mean of $f_i$.

\begin{align}
    SD &= \sqrt{\frac{1}{N} \sum^N_{i=1}(f_i - \bar{f})^2} \\
    SE &= \frac{SD}{\sqrt{N}} \label{eq:standardError}
\end{align}

\section{Experiment}
\begin{figure*}[bht]
    \centering
    \includegraphics[width=\linewidth]{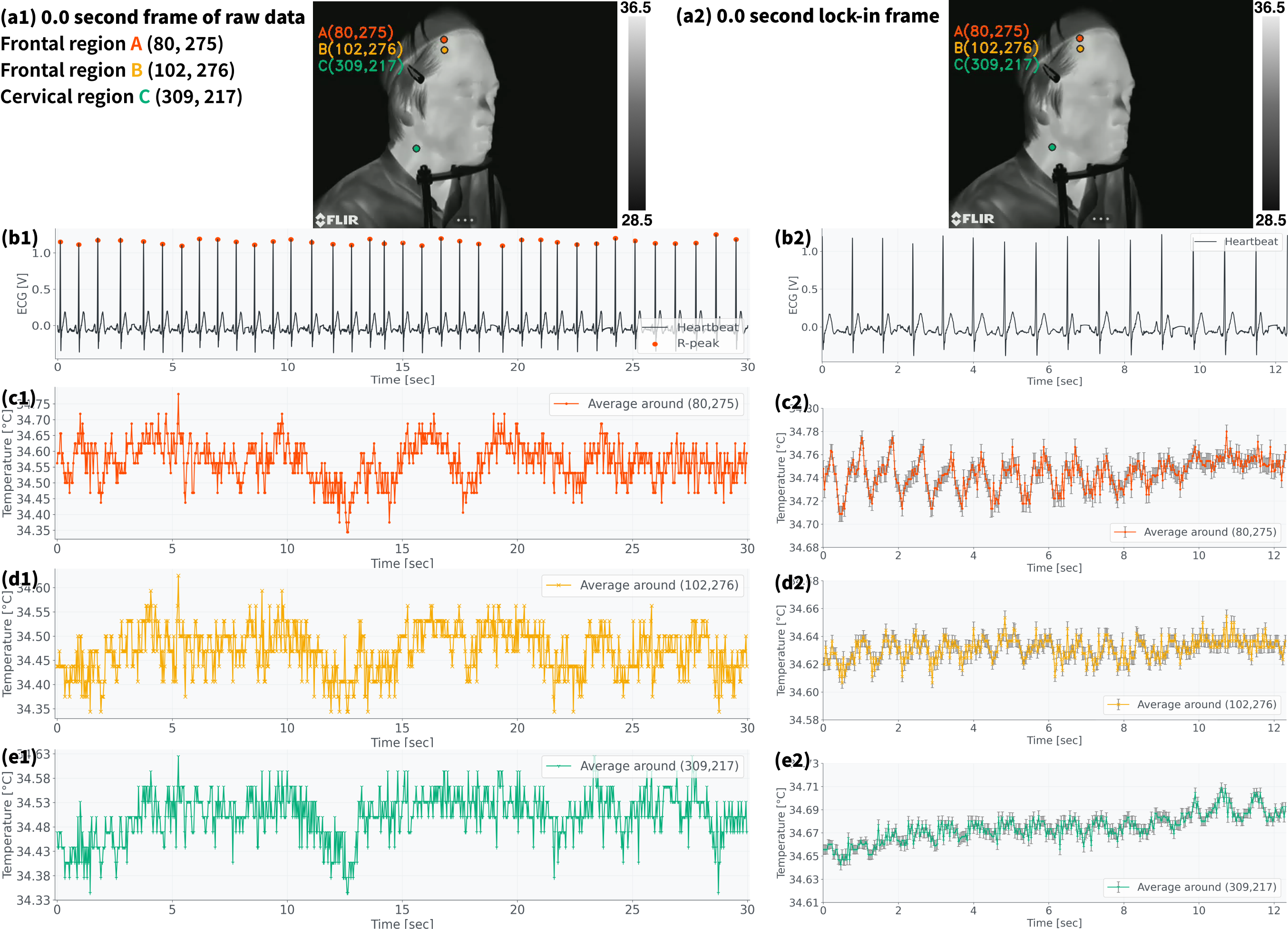}
    \caption{Comparison of raw data and lock-in thermography results for the male right profile. (a1) 0.0 second frame of raw data. (a2) 0.0 second frame of lock-in result. (b1) Heartbeat of raw data. (b2) Heartbeat normalized to mean RR interval at lock-in. (c1) Raw data of temperature at the frontal region A (80, 275). (c2) Lock-in thermography result at the frontal region A (80, 275). (d1) Raw data of temperature at the frontal region B (102, 276). (d2) Lock-in thermography result at the frontal region B (102, 276). (e1) Raw data of temperature at the cervical region C (309, 217). (e2) Lock-in thermography result at the cervical region C (309, 217).}
    \label{fig:ManRightSideFaceLock-inThermography}
\end{figure*}

\subsection{Participants}
Two healthy subjects, one male and one female, aged 20-30 years, participated in this study.
The series of experiments were conducted with the approval of the University's ethical review committee. The approval number is 23-466.
Prior to participating in the experiments, the study was explained to the participants, and written consent for participation was obtained.

\subsection{Equipment and Environment}
A thermal camera (FLIR T530) was used to acquire thermal images. FLIR T530 has a resolution of 480 $\times$ 640 pixels.
The thermal camera was positioned at 100 cm from the subject and 125 cm from the floor. In the meantime, an amplifier for biological signal measurement (AMP-151) was used to measure the ECG with electrodes attached to the subject. AMP-151 provides measurements amplified by a factor of 500.
The experiment was performed in a room maintained at room temperature \SI{25}{\degreeCelsius} and humidity of 10\%. 

\subsection{Experimental Procedure}
The experiment proceeded as follows.
\begin{enumerate}[leftmargin=14pt]
    \item  Subject Preparation: The subject receives a detailed explanation of the experiment and completes the consent form.
    \item  Equipment Setup: The subject attaches the ECG electrodes and positions his/her head in front of the thermal camera, ensuring that the hair and other obstructions do not cover the face.
    \item  Acclimatization: The subject rests for 10 minutes to acclimate to room temperature.
    \item  Initial Positioning: While seated, the subject places his/her head lightly on the chin rest and stabilizes it in a comfortable posture.
    \item  Data Collection -Front View: The subject remains seated in a resting position for 4 minutes. The heartbeat and face temperature are recorded. The face is captured from the front.
    \item  Data Collection -Right View: Procedure 5. is repeated with the face captured from the right side.
    \item  Data Collection -Left View: Procedure 5. is repeated with the face captured from the left side.
    \item  Break Interval: A 3-minute break between each recording session is given.
\end{enumerate}

\subsection{Signal Processing}
The proposed lock-in detection was implemented as a post-processing step.
Although recordings were made for 4 minutes each, only 30 seconds of the data with particularly low motion artifacts were used for analysis. Body motions were detected using optical flow and were also checked visually.

To extract a temporal temperature profile on a pixel of interest on a thermal image, the recorded temperatures are spatially averaged over 9 pixels, one central and 8 surrounding pixels, and then temporally averaged in the following manner. Initially, 30 seconds of the ECG data were analyzed to detect the R-peaks. Since the heartbeat is not perfectly periodic, the intervals between the adjacent R-peaks are non-uniform. To perform synchronous averaging of the face temperature using the R-peaks as a time stamp, the temporal axis of the R-peaks, hence the recorded temperature, is locally scaled so that the R-peaks appear periodically every second. $T_{RR}$ in \equref{eq:frameExtraction} from the average RR interval is calculated. Then, $N$ in \equref{eq:synchronousAdditiveAverage} is calculated from the lock-in window in this experiment as $N \cdot T_{RR}=15$ seconds. 
An average was then calculated based on \equref{eq:synchronousAdditiveAverage}. In this way, 15 seconds ($N \cdot T_{RR}$) of data were required to calculate one cycle of averaged temperature change. By sliding this lock-in window over the whole 30-second dataset, a temporal shift in the temperature change was investigated.  
In this experiment, the sliding interval was set to 1 second. 
The standard error is calculated from the number of locked-centered frames multiplied by 9 pixels based on \equref{eq:standardError}.

\subsection{Results}
An experimental demonstration of lock-in thermography for a male subject captured at the right view is summarized in \figref{fig:ManRightSideFaceLock-inThermography}. A snapshot of an original thermal camera image is shown in \figref{fig:ManRightSideFaceLock-inThermography} (a1), in which we particularly pick up three points A-C for explanation while data at other points can be processed in the same manner. 
The ECG and the three temperature profiles at A-C are recorded simultaneously as shown in \figref{fig:ManRightSideFaceLock-inThermography} (b1)-(e1). Using these data, we implement lock-in thermography \figref{fig:ManRightSideFaceLock-inThermography} (a2)-(e2), which can capture minute changes in the face temperature.
By detecting the R-peaks as marked in red in \figref{fig:ManRightSideFaceLock-inThermography} (b1), we locally scaled the temporal axis of the data as shown in \figref{fig:ManRightSideFaceLock-inThermography} (b2) to calculate the synchronous average over 16 cycles while sliding the window. The results at A-C are shown in \figref{fig:ManRightSideFaceLock-inThermography} (c2)-(e2), respectively. 
The error bars represent the standard deviation. 
In all cases, we observe periodic changes in the face temperature synchronous with the heartbeat in \figref{fig:ManRightSideFaceLock-inThermography}, suggesting a fluctuation caused by blood circulation in the frontal branch of the superficial temporal artery in (c2) and (d2) and the carotid artery in (e2). 
The waveform of temperature in \figref{fig:ManRightSideFaceLock-inThermography} (c2)-(e2) shows the same trend as that of the pulse generally resembles a triangular wave \cite{Vosse2011ARFM}.
It is noteworthy that the phase of periodic change at C is different from that at A;  A and ECG in (b2) are in quadrature phase, whereas C and ECG are in phase when referencing the R-peaks. This shows the potential of the proposed method to visualize the effect of different arteries. 

\begin{figure*}[tb]
    \centering
    \includegraphics[width=\linewidth]{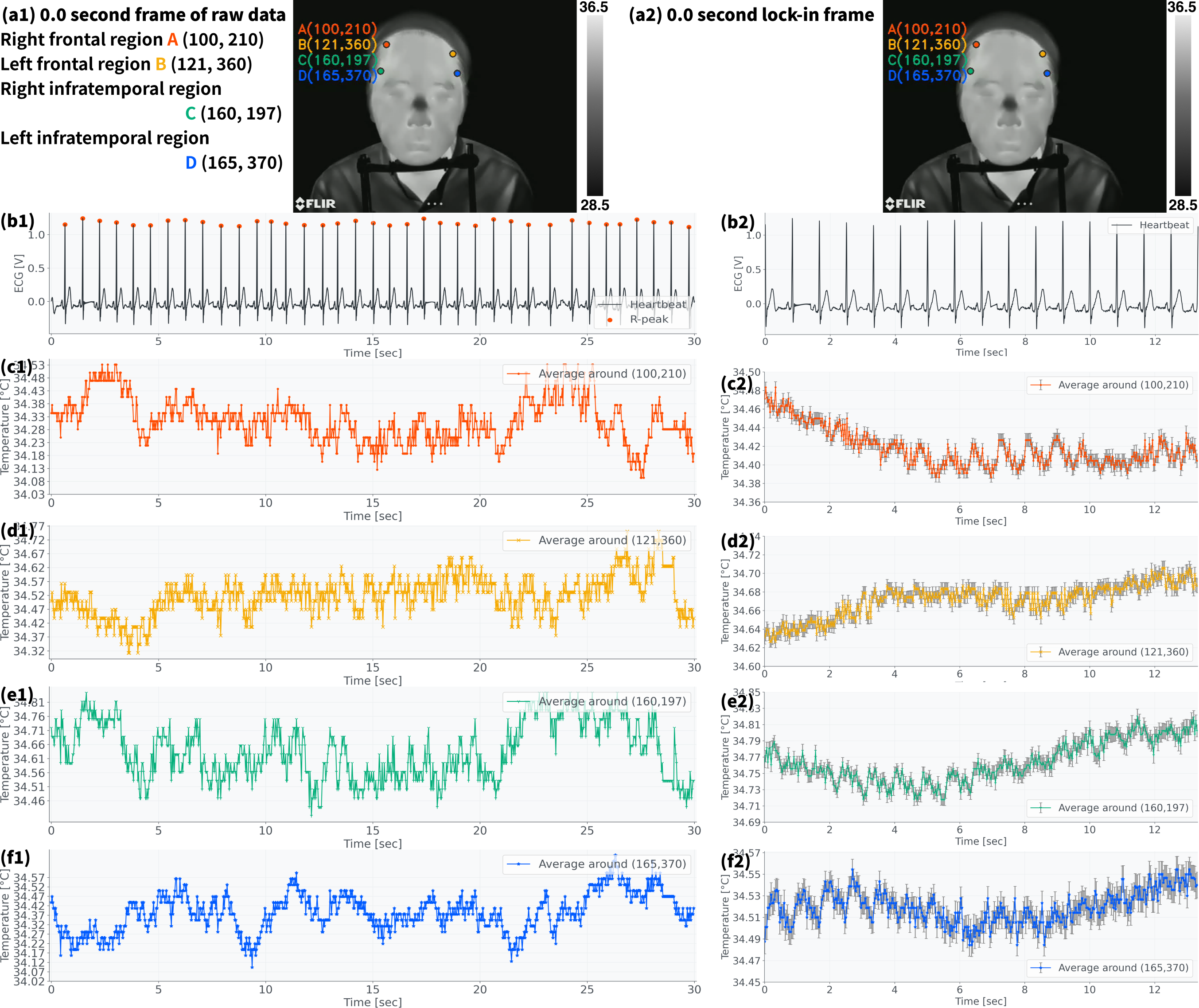}
    \caption{Comparison of raw data and lock-in thermography results for the male frontal face. (a1) 0.0 second frame of raw data. (a2) 0.0 second frame of lock-in result. (b1) Heartbeat of raw data. (b2) Heartbeat normalized to mean RR interval at lock-in. (c1) Raw data of temperature at the right frontal region A (100, 210). (c2) Lock-in thermography result at the right frontal region A (100, 210). (d1) Raw data of temperature at the left frontal region B (121, 360). (d2) Lock-in thermography result at the left frontal region B (121, 360). (e1) Raw data of temperature at the right infratemporal region C (160, 197). (e2) Lock-in thermography result at the right infratemporal region C (160, 197). (f1) Raw data of temperature at the left infratemporal region D (165, 370). (f2) Lock-in thermography result at the left infratemporal region D (165, 370).}
    \label{fig:ManFrontFaceLock-inThermography}
\end{figure*}
We next show results for the same subject as in \figref{fig:ManRightSideFaceLock-inThermography} in the front view in \figref{fig:ManFrontFaceLock-inThermography}. 
As expected, the original noisy data in \figref{fig:ManFrontFaceLock-inThermography} (c1) reveal sinusoidal temperature fluctuations after synchronous averaging around 4 to 10 seconds in \figref{fig:ManFrontFaceLock-inThermography} (c2). Such a sinusoidal change was relatively weak at region B on the left forehead, which is in a symmetrical position with A, as shown in \figref{fig:ManFrontFaceLock-inThermography} (d2). This is thought to be because, unlike the right frontal region, the left frontal region was shielded near the hairline, and thus the position of the left frontal region B could not be measured in perfect symmetry with the right frontal region A.
On the other hand, at infratemporal regions C and D in \figref{fig:ManFrontFaceLock-inThermography} (e2) and (f2), temperature changes that fluctuate in phase with the heartbeat referring to R-peak. At the right region C in (e2), the temperature rises at the timing of the R wave and falls between the R wave and the R wave. On the other hand, at left region D in (f1), the temperature falls at the timing of the R wave and rises between the R and R waves. At infratemporal regions C and D, temperature fluctuation is probably due to the influence of the superficial temporal artery.

\begin{figure*}[t!]
    \centering
    \includegraphics[width=\linewidth]{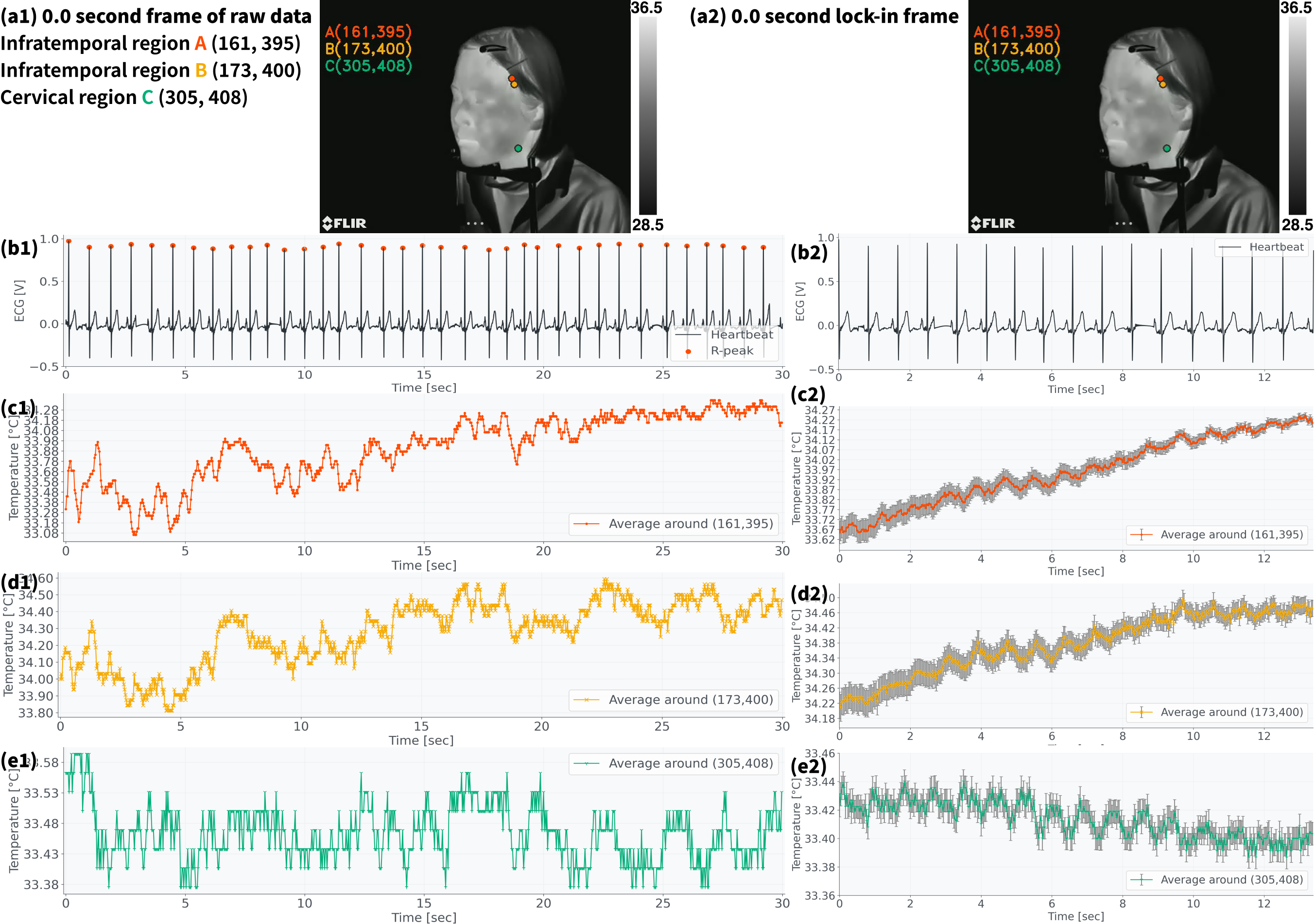}
    \caption{Comparison of raw data and lock-in thermography results for the female left profile. (a1) 0.0 second frame of raw data. (a2) 0.0 second frame of lock-in result. (b1) Heartbeat of raw data. (b2) Heartbeat normalized to mean RR interval at lock-in. (c1) Raw data of temperature at the frontal region A (161, 395). (c2) Lock-in thermography result at the frontal region A (161, 395). (d1) Raw data of temperature at the frontal region B (173, 400). (d2) Lock-in thermography result at the frontal region B (173, 400). (e1) Raw data of temperature at the cervical region C (305, 408). (e2) Lock-in thermography result at the cervical region C (305, 408).}
    \label{fig:WomanLeftSideFaceLock-inThermography}
\end{figure*}

We also confirm the repdocusiblity of the proposed method by observing face temperature changes with a female subject. 
The left view of the lock-in thermography is shown in 
\figref{fig:WomanLeftSideFaceLock-inThermography}(a2) to (e2). 
It can be seen that in the frontal regions A and B in (c2) and (d2), the face temperature repeatedly falls and rises synchronously with the R wave like a sinusoidal wave in the opposite phase. The temperature fluctuations in regions A and B can be attributed to the influence of the frontal branch of the superficial temporal artery. The result of \figref{fig:WomanLeftSideFaceLock-inThermography} (e2), which is locked in based on (e1), confirm periodic temperature fluctuations probably due to the facial artery.

For the frontal branch of the superficial temporal artery, the results in \figref{fig:ManRightSideFaceLock-inThermography} (c2) and (d2) and \figref{fig:ManFrontFaceLock-inThermography} (c2) were obtained from male subjects, and \figref{fig:WomanLeftSideFaceLock-inThermography} (c2) and (d2) from female subjects. The experiment showed the possibility that the proposed method can be applied regardless of the gender of the subjects.

\section{Discussion}
In this experiment, $N \cdot T_{RR}=15$ seconds were used as the lock-in window.
However, in future works, we will consider reducing $N$ to speed up the measurement. A possible trade-off is anticipated as follows.
If the lock-in window is too short, it is difficult to achieve a sufficiently high signal-to-noise ratio. On the other hand, if it is too long, the effects of body motion artifacts and long-term drifts reduce the coherence during the window.
Considering these trade-offs, we plan to qualitatively and quantitatively study the condition for an effective implementation of lock-in thermography. 

While we have observed sinusoidal periodic change of the face temperature with lock-in thermography, the clarity of the oscillatory behavior seems time-dependent. This is simply explained by the body motion of the subjects, which shifts the position of the arteries. Therefore, we need to implement tracking of the arteries by analyzing all the pixels simultaneously, instead of selecting a few discrete observation points. By calculating Fourier transformation at each pixel and searching points with the largest oscillation amplitude at the same interval as the heartbeat cycle, we can track pixels with clear temperature changes corresponding to the arteries. 
This will be useful for automatically detecting ROI suited for emotion estimation in future works.

\section{Conclusion}
We have proposed and demonstrated lock-in thermography using the ECG as a reference signal. 
The minute temperature changes, less than 10 mK, can be detected with the proposed method owing to the improved signal-to-noise ratio.

While heartbeat was measured with ECG by wearing electrodes on the skin in this study, other simpler tools such as smartwatches can also be used to detect the R-peaks. 
Since research on non-contact measurement of heartbeat is also actively conducted \cite{Matsumoto2020NE}, it is possible to measure the heartbeat in a completely non-contact manner in combination with thermal imaging cameras.
We have used the heartbeat measured in 15 seconds for synchronous averaging. Reducing the number of averaging involves trade-offs between accuracy and speed. 
It is important to utilize the signal waveform of the heartbeat beyond the R-peaks to improve the signal-to-noise ratio without increasing the acquisition time.

After establishing lock-in thermography via the implementation of the above improvements, we will tackle the challenge of emotion estimation based on appropriate ROI selection. Many studies on emotion estimation using infrared thermography have reported inconsistent results due to differences in experimental settings. Using lock-in thermography, with which we can identify the location of the arteries, we aim to absorb noise and environmental differences and achieve stable emotion estimation. The simple setup of lock-in thermography allows us to apply the proposed method not only in a laboratory but also in more dynamic and practical situations.

\section*{Acknowledgements}
This work was supported by JSPS Grant-in-Aid for Challenging Research (Pioneering)  Grant Numbers JP21K18307 and Grant-in-Aid for Scientific Research (A) Grant Number JP24H00704.

\bibliographystyle{ieeetr}
\bibliography{ref}




\end{document}